\documentclass[aps,prb,showpacs,twocolumn,superscriptaddress]{revtex4}
\usepackage{epsfig}
\usepackage{amsfonts}
\usepackage{amsmath}
\usepackage[T1]{fontenc}

\newcommand{\ket}[1]{|#1\rangle}

\newcommand{\beq}{\begin{equation}}
\newcommand{\eeq}{\end{equation}}
\newcommand{\bea}{\begin{eqnarray}}
\newcommand{\eea}{\end{eqnarray}}
\newcommand{\nn}{\nonumber}

\newcommand{\al}{\alpha}
\newcommand{\s}{\sigma}

\newcommand{\p}{\partial}

\newcommand{\ua}{\uparrow}
\newcommand{\da}{\downarrow}

\newcommand{\dmi}{\frac{1}{2}}
\newcommand{\lra}{\longrightarrow}

\begin{document}
\title{Incompressible states of a two-component Fermi gas in a double-well optical lattice}
\author{Fran\c cois Cr\'epin}
\affiliation{Laboratoire de Physique des Solides, CNRS UMR-8502 Universit\'e Paris Sud, 91405 Orsay Cedex, France}
\author{ Roberta Citro}
\affiliation{Dipartimento di Fisica "E.R. Caianiello"- Universita di Salerno Via S. Allende I-84081 Baronissi (Sa) - Italy}
\author{Pascal Simon}
\affiliation{Laboratoire de Physique des Solides, CNRS UMR-8502 Universit\'e Paris Sud, 91405 Orsay Cedex, France}

\date{\today}


\begin{abstract}
We propose a scheme to investigate the effect of frustration on
the magnetic phase transitions of cold atoms confined in an
optical lattice. We also demonstrate how to get two-leg spin
ladders with frustrated spin-exchange coupling which display a
phase transition from a spin liquid to a fully incompressible
state. Various experimental quantities are further analyzed for
describing this new phase.
\end{abstract}


\maketitle

\section{Introduction}

Quantum simulation has been put forward as a tool to probe the
physics of a variety of many-body systems. In particular, schemes
with cold atoms in optical lattices have been suggested to
simulate arbitrary spin
models\cite{zoller_2003,svistunov_2003,demler_2003,cirac_spin_dynamics_2003,yip_2003,
cirac_spin_hamiltonians_2004} and some interesting frustrated
Hamiltonians.\cite{santos_kagomelattice_2004,polini_2005} Compared
to the alternative of looking for or inducing low-dimensional
behavior on existing magnetic
materials\cite{gopalan_1994,kitaoka_1994} and in organic
conductors,\cite{organic_ladder} cold atoms and molecules
offer greater flexibility in terms of variable geometry and
interaction strength.

By adjusting the amplitudes and propagation directions of laser
beams, a variety of geometries of optical lattice are generated.\cite{blakie,petsas} This technique offers the possibility for constructing
spin systems, including spin chains,
\cite{cirac_spin_hamiltonians_2004,9,demler_2003} kagome lattices,
\cite{santos_kagomelattice_2004} and spin ladders.
\cite{cirac_spin_hamiltonians_2004,12,13} However, the majority
of current proposals are perturbative and their effective
interactions are rather weak. This makes their experimental
realization challenging, as it requires very low temperatures. A
solution suggested to solve the problem of weak interactions is to
replace the atoms with polar
molecules.\cite{zoller_polar_molecules}

In experiments, in addition to cold Bose atoms, Fermi atoms in optical
lattices have also been realized recently, such as $^{40}$K in
one-dimensional (1D) \cite{14} and 3D  lattices\cite{15}, $^6$Li
in 3D lattices,\cite{16} and a variety of interesting phenomena
as for example, Bloch oscillations, band insulators, and
superfluidity were reported.


One fundamental issue in magnetism is to understand how frustration affects
the long range properties of spin-models.
Frustrated models are described by Hamiltonians with
competing local interactions such that the system cannot
minimize its energy so as to satisfy these constrains simultaneously.
Frustrated models typically
have highly degenerate ground states, which can become ordered by
increasing the temperature or by quantum fluctuations,  i.e.
 ``order by disorder''. Theoretical and numerical issues, such as
the large dimensionality or the sign problem in Monte Carlo
simulations make it very difficult to study frustrated
Hamiltonians, while cold atoms in optical lattices may provide an
alternative tool to analyze the effects of frustration.

In this paper, we propose an optical lattice setup to produce
quasi-1D frustrated spin ladders to illustrate a quantum phase
transition from a spin-liquid phase to a fully incompressible
state (FIS). Two-component ultracold fermionic atoms loaded in ladder-like optical lattices have also been
proposed recently as good candidates to observe and study
Fulde-Ferrell-Larkin-Ovchnnikov (FFLO) modulating superconducting instabilities \cite{feiguin09,fujihara}
Such  FIS is in correspondence with a magnetization plateau at an intermediate magnetization.
It occurs at half-filling and should be observed in the center of the trap. We
focus on experimental observables able to distinguish between such phases.


The plan of the paper is the following: In Sec. II, we show how to
generate ladder-like structures with four pairs of lasers. In Sec.
III, we derive and study the resulting effective spin model at
half filling in the limit of a strong inter-chain coupling. Sec.
IV is devoted to the analysis of various observables  in order to
characterize the phases with both a charge and ``spin'' gap, the
so-called fully incompressible phase. Finally, in Sec. V, we
summarize our results and formulate some perspectives.



\section{Ladder like structures with cold atoms}\label{sec:2}
\begin{figure}
\epsfig{figure=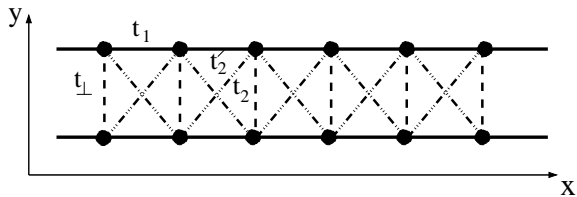,width=7.6cm}
\caption{A schematic view of the frustrated two-leg ladder. Each black dot represents a well of the potential.
The atoms can tunnel between neighboring wells along the chains (plane line), along the rungs (dashed lines)
and also along   the two diagonals (dot-dashed lines) with respective amplitudes $t_1,t_\perp$ and $t_2,t'_2$.}\label{Fig:ladder1}
\end{figure}

In this section, we describe how to obtain a ladder like structure with laser beams.
The typical $2-$leg ladder we have in mind is depicted in Fig. \ref{Fig:ladder1}.

\begin{figure}
\epsfig{figure=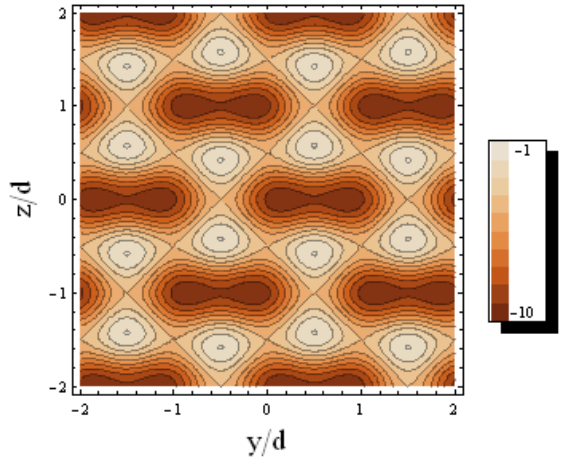,width=7.6cm}
\caption{(Color online) Example of a double-well lattice potential in the $(y,z)$ plane. Distances are in units of the lattice spacing $d=\lambda/2$. The energy unit is $E_R$ (recoil energy). The $x$ direction corresponds to the ladder axes.}\label{Fig:array2}
\end{figure}

In order to obtain ladder-like structures from an optical lattice,
the strategy is to obtain first a lattice of double-wells. This
can be realized by superimposing two independent planar optical
lattices with periodicity $\lambda$ and $\lambda/2$. In the
following we denote $d=\lambda/2$ the typical lattice spacing and
set $d=1$. The  energy scale of the atoms in the lattice is
determined by the recoil energy $E_R = \hbar^2 k^2 /2m$. Such an
array of double-wells has been proposed in [\onlinecite{porto06}]
and realized in [\onlinecite{porto07}] with red-detuned lasers.
Following Ref. [\onlinecite{porto06}], the resulting 2D potential
is given by: \bea
V_p(y,z)&=&-\big[2\cos(2\pi y)+2\cos(2\pi z)+4\big]\\
&-& 16R\cos^2(\pi\frac{y-z}{2}-\frac{\pi}{4})\cos^2(\pi
\frac{y+z}{2}-\frac{\pi}{4}),\nn \eea where the energy unit is the
recoil energy, $R$ is an adjustable parameter related to the
relative difference of intensity between the two lattices. By
tuning $R$, one obtains an array of double-wells, an example of
which is represented in Fig. \ref{Fig:array2}. By adding an
independent standing wave in the $x$ direction, one therefore
obtains a 3D lattice of ladders.

\begin{figure}
\epsfig{figure=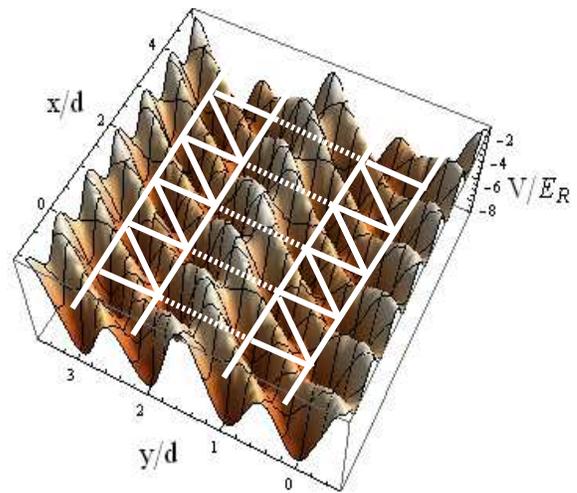,width=7.6cm}
\caption{(Color online) Example of a  potential in the $(x,y)$ plane leading to a frustrated ladder-like potential.
The frustrated ladders are sketched on the potential.}\label{Fig:ladder2}
\end{figure}

This kind of potential is suitable to obtain a lattice of
non-frustrated ladders. Indeed, since the potential along (x,y) is
separable, and Wannier functions at different sites are
orthogonal,  the diagonal tunneling matrix elements between wells
are zero. In order to have a frustrated ladder as depicted in Fig.
\ref{Fig:ladder1}, a non-separable potential along (x,y) is
required. It can be achieved by using two interfering standing
waves with a blue-detuned laser along the diagonals, in the (x,y)
plane. Such a potential is given for instance by: \beq V_d(x,y) =
\big[ \sqrt{V_+} \sin(\pi(x+y)) + \sqrt{V_-} \sin(\pi(x-y))
\big]^2. \eeq Note that for this additional potential to be
independent from the other, the laser frequency must be 
detuned with respect to the in-plane lasers, since all three orthogonal polarizations are already used to create the ladders. In practice all lasers are detuned in order to supress any residual interference. For simplicity, we
use the same lattice spacing in the $x$ direction and thus the total
optical potential is given by \beq \label{eq:potential} V(x,y,z) =
V_p(y,z) + V_d(x,y), \eeq It describes a 3D array of frustrated
2-leg ladders.
 An example of the total potential is represented in Fig. \ref{Fig:ladder2}.

\section{Model Hamiltonian}

In the following, we assume that hopping processes and
interactions leave the atoms in the lowest band. We assume that
the trap is loaded with fermionic species with two different
hyperfine ``spin'' states. In what follows, we denote these states
by $\s=\ua,\da$ by analogy with a spin $1/2$. From the potential
written in Eq.(\ref{eq:potential}), we derive an Hubbard type
Hamiltonian:

\bea
\label{hubbard}
H_{Hub} &=& -\sum_{i,\alpha=1,2,\s} t_1(c^\dag_{i,\alpha,\s} c_{i+1,\alpha,\s}+h.c.)\nn \\
&-&\sum_{i,\s}( t_\perp c^\dag_{i,1,\s} c_{i,2,\s} + t_2  c^\dag_{i,1,\s} c_{i+1,2,\s}\nn \\
&&~~~~ + t'_2 c^\dag_{i,2\s} c_{i+1,1\s}+h.c.)\nn \\
&+&  \sum_{i,\alpha}(U n_{i,\alpha,\ua} n_{i,\al, \da}-\mu_{i,\ua}
n_{i,\al,\ua}-\mu_{i\da}n_{i,\al,\da}). \eea In this Hamiltonian,
$c_{i,\s,\alpha}$ destroys an atom with ``spin'' index  $\s$ on
site $i$ of the chain $\alpha$,
$n_{i,\al,\s}=c^\dag_{i,\alpha,\s}c_{i,\alpha,\s}$ is the
occupation at site $i$ of the chain $\alpha$.
$t_1,t_\perp,t_2,t'_2$ are tunneling amplitudes as defined in Fig.
\ref{Fig:ladder1}, $U$ is a local interaction which can be
controlled by Feshbach resonances and $\mu_{i\s} = \mu_\s - V_{i,\s}$ is a local
chemical potential. $V_{i,\s}$ is the harmonic confining potential in the $x$
direction, while $\mu_\s$ is a global chemical potential for species $\s$. Although in real experiments the number of atoms is fixed and constant we work here in the grand-canonical ensemble, with a fixed chemical potential, in order to make a connection with magnetic systems. We assume for now that the trap is sufficiently elongated in
the $x$ direction to take a constant $\mu_{i\sigma}=\mu_\s$.
Then, the last term
in Eq. (\ref{hubbard}) can be rewritten as
$$-\mu\sum_{i,\alpha} ( n_{i,\al,\ua}+ n_{i,\al,\da})-h\sum_{i,\alpha}( n_{i,\al,\ua}- n_{i,\al,\da}),$$
with $\mu=(\mu_{\ua}+\mu_\da)/2$ and
\beq\label{eq:heff}
h=(\mu_{\ua}-\mu_\da)/2,
\eeq
 which are respectively equivalent to a chemical potential and
an magnetic field. In what follows, when we will mention
the magnetic field, we refer to an effective magnetive field given by Eq. (\ref{eq:heff}).
We focus on the case of strong repulsive interactions ($U \gg
t_\perp,t_1,t_2,t'_2$) and assume  half-filling, i.e. $\langle n_{i,\ua}+n_{i,\da}\rangle = 1$ in a homogeneous trap. In this limit,
the system is in a Mott insulator phase with one atom per site.
Charge degrees of freedom are frozen and only superexchange
interactions are allowed to second order. These virtual processes
lead to antiferromagnetic couplings between sites. After projecting on states with exactly one atom per site, the effective
Hamiltonian for a single ladder reads:
 \bea \label{Ham_lad}
H_{lad} &=& J_\perp \sum\limits_{i=1} {\bf S}_{i,1} {\bf
S}_{i,2}+
J_1 \sum\limits_{i=1} \sum\limits_{\al=1}^2{\bf S}_{i,\al} {\bf S}_{i+1,\al}\nn\\
&+&J_2 \sum\limits_{i=1} {\bf S}_{i,1} {\bf S}_{i+1,2}+
J'_2 \sum\limits_{i=1} {\bf S}_{i,2} {\bf S}_{i+1,1}\nn\\
&-&h\sum\limits_{i=1} \sum\limits_{\al=1}^2 S_{i,\alpha}^z, \eea
where ${\bf S}_{i,\alpha}=\sum_{\s\s'}
c^\dag_{i,\alpha,\s}\frac{\vec \tau_{\s\s'}}{2} c_{i,\alpha,\s'}$
is the ``spin'' operator, and $\vec \tau$ are the Pauli matrices. In
Eq. (\ref{Ham_lad}), $J_i = 4t_i^2/U >0$ denotes a superexchange
coupling. We compare relative spin exchange couplings by estimating the tunneling matrix elements. We approximated the
Wannier functions with the ground state of the harmonic
oscillator\cite{review08} in the bottom of each well. With
$R=0.14, V_+ = 0.1 E_R$,  $V_- = 4 E_R$, we found:
$(t_\perp/t_1)^2 = 20,\  (t_2/t_1)^2 = 1.5,\ (t'_2/t_2)^2 =
10^{-5}.$ We also estimated inter-ladder couplings and found $\ (t_{2,\textrm{inter}}/t_2)^2 = 6.10^{-3},\
(t_{1,\textrm{inter}}/t_1)^2 = 9.10^{-3}$.
>From these estimations we can infer that the ladders are very
weakly coupled together\cite{note}. Moreover, when the temperature
is much larger than the interladder exchange energy, i.e. $k_BT\gg
J_{3D}$, the ladders are effectively decoupled. In this limit, we
can thus consider a single ladder-like lattice potential, as described by Hamiltonian (\ref{Ham_lad}).

Although we assume that the system is at half-filling {\it i.e.}
that the average density per site  $\rho=\rho_\ua+\rho_\da =1$,
the densities $\rho_\ua$ and $\rho_\da$ can be quite different. In
this respect, $h$ is related to the difference between the
chemical potentials of both fermionic species (see Eq. (\ref{eq:heff})).
The Hamiltonian thus reduces to that of an Heisenberg ladder in a magnetic
field $h$. Such an Hamiltonian has been studied by various methods
(e.g. see \cite{cabra97,chitra97,totsuka,mila98,chaboussant98,chitra99}).



Let us note that in real systems, an additional weak isotropic
harmonic potential exists over the lattice\cite{schneider_04}. This
confinement brings an energy offset on each lattice site and leads
to an effective local chemical potential. The superfluid-insulator
phase diagram is influenced by this potential. In the case where the trapping potential is spin-independent, there is no influence on the pseudo magnetic field $h$. If the interaction energy $U$ is much larger than the trapping energy $E_{\textrm{trap}} = \frac{1}{2} m \omega_{x}^2 N_\sigma^2$, with $\omega_x$ the trapping frequency in the ladder direction and $N_\sigma$ the number of $\sigma$ atoms, one can expect a Mott insulator phase in the center of the trap, surrounded by a superfluid region.

\section{Spin and charge phases}\label{sec:4}

For our analysis let us start from the strong coupling limit
$J_\perp\gg J_1,J_2,J'_2$. This limit is clearly met within our
proposed optical lattice potential.

When $J_1 = J_2=J'_2= 0$ each rung is independent, and has four available states, a singlet and three triplets:
 $\ket{S} =\frac{1}{\sqrt{2}}(\ket{\ua \da}-\ket{\da \ua})$, $\ket{T_{-1}} = \ket{\da \da}$, $\ket{T_{0}} =  \frac{1}{\sqrt{2}}(\ket{\ua \da}+\ket{\da \ua})$ and $\ket{T_{1}} = \ket{\ua \ua}$. Their energies are $E(S) = -3J_\perp/4$, $E(T_{1}) =J_\perp/4 - h$, $E(T_{0}) = J_\perp/4$ and $E(T_{-1}) = J_\perp/4 + h$ respectively. \\
Upon increasing the effective magnetic field, the ground state
rung magnetization $M = \langle S^z_{i,1} \rangle + \langle
S^z_{i,2} \rangle$ increases up to saturation at $h = h_c =
J_\perp$ where $\ket{S}$ and $\ket{T_1}$ become degenerate.  In
the presence of an inter-rung interaction the transition becomes
broader and a crossover is expected between two critical fields $h_{c1}$ and $h_{c2}$.

\subsection{Special case $\rho_\ua=\rho_\da=1/2$}
\subsubsection{Repulsive interactions}
In the case of half-filling with equal densities
$\rho_\ua=\rho_\da=1/2$, the strong coupling picture predicts a
spin gap at zero magnetization, $M=0$. This state corresponds to a
Fully Incompressible State (FIS) where both charge and spin
degrees of freedom are gapped. We point out that this
incompressible phase does not require non-zero couplings
$J_2,J'_2$ or equivalently non-zero tunneling terms $t_2,t'_2$.
Consequently, one can reach this phase with $V_d=0$ in the
potential in Eq. (\ref{eq:potential}). In the next subsection, we
therefore assume $t_2=t'_2=0$.

Let us note that the spin gap is very robust and can be
theoretically obtained  also in the weak coupling $J_\perp\ll J_1$
using a bosonization approach.\cite{giamarchi04}

\subsubsection{Attractive interactions}
The spin Hamiltonian in Eq.(\ref{hubbard}) was derived assuming
strong repulsive interactions $U\gg t>0$. When $U<0$,
states with zero or two atoms per site are favored. Therefore the
low-energy excitations are the ones that describe the dynamics of
atom pairs (in the singlet state) in a periodic potential. One may
wonder whether the FIS survives at strong attractive interactions.

The correspondence is in fact quite general and can be obtained by
performing the following particle-hole transformation on the
initial Hubbard Hamiltonian in Eq. (\ref{hubbard}): \bea\label{ph}
c_{i,\alpha,\ua}&\lra& c_{i,\alpha,\ua}\nn\\
c_{i,\alpha,\da}&\lra& (-1)^{i+\al} c_{i,\alpha,\da} \eea This
transformation affects only the ``spin'' $\da$ atoms. Provided
$t_2=t'_2=0$, the tunneling part of the Hamiltonian is left
unchanged. However, $n_{i,\al,\da}\to (1- n_{i,\al,\da})$. This
implies therefore that $U\to -U$ in Eq. (\ref{hubbard}). The
particle-hole transformation exchange the roles of $\mu$
and $h$, up to unimportant constants. Therefore, the fully gapped
phase with $\rho_\ua=\rho_\da=1/2$ that we described above from
positive $U$ maps into another equivalent gapped phase at negative
$U$.

Using this correspondence, one can easily infer the evolution of
the average number of pairs on a rung as function of the chemical
potential for $U<0$ simply by looking at the evolution of the
``magnetization'' with respect to the magnetic field $h$, for $U>0$. In fact half-filling corresponds to
one pair of atoms per rung  and to $\mu=0$. In order to increase
the number of atoms beyond half-filling, one has to therefore to
fill a gap of order $J_\perp$.

\subsection{The generic case $\rho_\ua\ne\rho_\da$}
\subsubsection{Repulsive interactions}
In order to study the physics at $M\ne 0$, we follow
Ref. [\onlinecite{mila98}] and give the main steps of the calculations leading to an effective 1D bosonic hamiltonian given in Eq. (\ref{eq:hbos}) for completness and readability. We first  separate the Hamiltonian $H_{lad}$
into two parts, $H_{lad}=\mathcal{H}_0 + \mathcal{H}_1$ with

\bea
\mathcal{H}_0 &=& J_\perp \sum\limits_{i=1}^N {\bf S}_{i,1} {\bf S}_{i,2} - h_c\sum\limits_{i=1}^N \sum\limits_{\al=1}^2 S_{i,\alpha}^z\\
\mathcal{H}_1 &=& J_1 \sum\limits_{i=1}^N \sum\limits_{\al=1}^2{\bf S}_{i,\al} {\bf S}_{i+1,\al}+
J_2 \sum\limits_{i=1}^N {\bf S}_{i,1} {\bf S}_{i+1,2}\nn\\
&+&
J'_2 \sum\limits_{i=1}^N {\bf S}_{i,2} {\bf S}_{i+1,1}-(h-h_c)\sum\limits_{i=1}^N \sum\limits_{\al=1}^2 S_{i,\alpha}^z
\eea

\bigskip
The ground-state of $\mathcal{H}_0$ is $2^N$ degenerate. The
perturbation Hamiltonian $\mathcal{H}_1$ will lift the degeneracy.
We use standard quantum mechanics perturbation theory for
degenerate ground  states. Diagonalization of the perturbation
Hamiltonian in the degenerate subspace yields the result to first
order. One can rewrite the effective perturbative Hamiltonian for
the system as:

\beq \label{hxxz}
\mathcal{H}_{eff} = \sum_{i=1}^N \left[ J_{xy} \left( \sigma^x_i \sigma^x_{i+1} + \sigma^y_i \sigma^y_{i+1}\right) + J_z\sigma^z_i \sigma^z_{i+1}\right] - \tilde h\sum_{i=1}^N\sigma_i^z
\eeq

with:
\bea
\label{Spin_ladd_to_XXZ}
&S_{i,1}^+=-\frac{1}{\sqrt 2}\sigma_i^+~, ~S_{i,2}^+=\frac{1}{\sqrt 2}\sigma_i^+\nn\\
&S_{i,1}^-=-\frac{1}{\sqrt 2}\sigma_i^-~, ~S_{i,2}^-=\frac{1}{\sqrt 2}\sigma_i^-\nn\\
&S_{i,1}^z =S_{i,2}^z=\dmi(\sigma_i^z+\dmi)~ ~, \eea

\noindent and $J_{xy}=J_1-(J_2+J'_2)/2$, $J_z=J_1/2+(J_2+J'_2)/4$,
and $\tilde h=h-h_c-J_1/2-(J_2+J'_2)/4$. The sigmas stand for
pseudo-spin 1/2 degrees of freedom. Formally these operators are
the Pauli matrices, written in the basis $\{\ket{\ua}_i =
\ket{T_1}_i, \ \ket{\da}_i = \ket{S}_i\}$. Therefore one is able
to map the original ladder system into a 1D Hamiltonian, namely
the so called $XXZ$ chain. Then one can
show (see for example \cite{giamarchi04}) using a Jordan-Wigner transformation that
the XXZ chain is equivalent to a single band of interacting
spinless fermions.  This correspondence allows us to describe
low-energy properties of the XXZ chain through bosonization
\cite{giamarchi04} and obtain correlation functions in this
regime. Using usual bosonization formulas one finds:
\beq
\label{sigma_z}
\sigma^z(x) = m - \frac{1}{\pi}\p_x\phi(x) +\frac{1}{2\pi \alpha} 2 \cos \left[2\phi(x) - \pi(2m+1)x\right],
\eeq

\beq
\label{sigma_+}
\sigma^+(x) =  \frac{(-1)^x}{\sqrt{2\pi\alpha}}e^{-i\theta(x)}\left[ 1 + \cos \left(2\phi(x) - \pi(2m +1)x  \right) \right].
\eeq
\noindent where the bosonic
fields $\phi$ and $\theta$ satisfy the commutation relation
$[\phi(x),\nabla \theta(x')]=i \pi \delta(x-x')$. One can look at these fields as the angles of a spin $1/2$ on the Bloch sphere, with $2\phi$ being the angle taken from the $z$ axis while $\theta$ is the angle in the $XY$ plane. The commutation relation implies that order in one direction will destroy order in the other direction. Note also that $m$ is the magnetization of the $XXZ$ chain and is related to the ladder magnetization through $M = \langle S_{i,1}^z + S_{i,2}^z \rangle = m + 1/2$. The form of the Hamiltonian now depends on the value of $m$. If $m \neq 0$ ($M > 1/2$), it simply reads:
\beq\label{eq:hbos}
H = \frac{1}{2\pi}\int dx \left[ uK\left( \partial_x \theta(x)\right)^2 + \frac{u}{K} \left( \partial_x \phi(x)\right)^2 \right]
\eeq

with
\bea
uK &=& v_F = J_{xy} \sin(k_F )\label{Lutt_param1}\\
\frac{u}{K} &=& v_F\left( 1 + \frac{2 J_z}{\pi v_F}(1-\cos(2k_F))\right)  \label{Lutt_param2}\\
\label{Lutt_param3} \eea

\noindent $u, K$ are the
so-called Luttinger parameters and control all low-energy
properties of the system. Their expression in Eqs.
(\ref{Lutt_param1}-\ref{Lutt_param3}) is obtained in the limit
of small $J_z$.

The simple quadratic form leads to a power-law decay of all correlation functions. The slowest decaying mode indicates quasi long-range order. The situation is quite different when $m=0$ ($M=1/2$). This corresponds to
$\rho_\ua=3/4$ and $\rho_\da=1/4$ in our initial fermi-fermi
mixture and when the ladder-like optical lattice is loaded with this
particular filling, the ``spin'' sector of the mixture is
described by a sine-Gordon model:
\bea
\label{SF_cosine}
H &=& \frac{1}{2\pi}\int dx \left[ uK\left( \partial_x \theta(x)\right)^2 + \frac{u}{K} \left( \partial_x \phi(x)\right)^2 \right] \nn \\
&-& \frac{2g_3}{(2\pi \alpha)^2} \int dx \cos \left(4 \phi(x)\right),
\eea
where $g_3 =  J_z$. 
This signals a possible tendency towards ordering in the $z$ direction. Energetically, it is favorable to lock the angle $\phi$ so as to minimize the cosine term. However it would create large fluctuations in the $\theta$ angle therefore costing a great amount of kinetic energy. This competition can therefore lead to a phase transition at this particular filling. A renormalization group calculation will indicate whether or not the cosine term is relevant at low energy. If yes, it will favor ordering in the $z$ direction and open a spin gap. The flow equations are:
\beq \frac{dK}{dl} = -y^2(l)K^2(l), \ \ \frac{dy}{dl} = (2-4K(l))
y(l). \eeq \noindent where we have defined  $y=g_3/(\pi u)$. $(K=1/2, y=0)$ is a fixed point. For $K>1/2$,
$dy/dl<0$ and the cosine is irrelevant. The system is
therefore a Luttinger liquid, even when $M=1/2$. However, for
$K<1/2$ the cosine is relevant and $y$ flows to strong coupling. A gap opens in the excitation
spectrum. From the Bethe ansatz solution of the XXZ chain, one can
infer that the value $K=1/2$ corresponds to the
isotropic case $J_z=J_{xy}$ \cite{giamarchi04}. The condition
$K<1/2$ is associated to the Ising phase of the XXZ Heisenberg
chain {\it i.e} $J_z>J_{xy}$. For our original ladder system, this
phase is realized when \beq \label{frustration}
J_2+J'_2>2J_1/3.\eeq
 Therefore, the opening of a
``spin'' gap can only occur for strong frustration.\cite{totsuka,mila98}

In what follows,  we consider two cases: (i) a regular ladder with
$J_2 = J'_2=0$ and (ii) a frustrated ladder with $J_2+J'_2
>2J_1/3$. In particular, as we have seen in Sec. \ref{sec:2}
frustration can be generated in the ladder-like optical lattice
with interfering standing waves along one diagonal direction (i.e.
when $J_2=0$ or $J'_2=0$).

For the weakly frustrated ladders (first case), we therefore
expect ``spin'' correlations functions to be of Luttinger-liquid
type when $\rho_\ua\ne \rho_\da\ne 1$ while for strongly
frustrated ladders (second case) we expect a ``spin'' gap opening
for the special filling $\rho_\ua=3/4,\rho_\da=1/4$. This
situation in real space would correspond to an alternation of
singlets and triplets.
Such a fully incompressible state would be a direct consequence of
frustration. In next section \ref{sec:observables}, we will
discuss how to distinguish these two phases by  computing various
observable.

\subsubsection{Attractive interactions}
An interesting question regards the survival of the
frustration-induced FIS when attractive interactions are
considered. Let us then apply the particle-hole transformation
defined in Eq. (\ref{ph}). This transformation makes the tunneling
terms along the diagonals of the ladder ``spin''-dependent. More
specifically, this changes the sign of the amplitudes $t_2$ and
$t'_2$ in Eq. (\ref{hubbard}) for the ``spin'' $\da$ atoms only. There is no direct mapping between the repulsive and the attractive cases. However, one can follow a procedure similar to the one used for the repulsive
case -- namely a Schrieffer-Wolff transformation -- starting directly from a Hamiltonian with attractive interactions, and derive a ``pseudo-spin'' ladder Hamiltonian quite
similar to Eq. (\ref{Ham_lad}) but breaking the $SU(2)$ particle-hole symmetry. After
projecting onto a XXZ ``pseudo-spin'' chain Hamiltonian similar to
Eq. (\ref{hxxz}), the coupling $J_{xy}$ now becomes
$J_{xy}=J_1+(J_2+J'_2)/2$. The condition $J_z>J_{xy}$ is no longer
met and thus the FIS found for positive $U$ does not survive for
negative $U$ in the case of a frustrated ladder.

\section{Observables}\label{sec:observables}
\subsection{Density correlations.}
After turning off the trap, and assuming that interactions are small, the atomic cloud evolves freely. It is possible,
with laser imaging, to measure both the density $\rho(r)$ defined by
\beq \rho_t(r)=\langle n(\mathbf{r}) \rangle_t\eeq
 and the density
correlation function $\mathcal G$ defined by
\beq \label{defG}
\mathcal{G}(\mathbf{r} - \mathbf{r}') = \langle n(\mathbf{r}) n(
\mathbf{r}')\rangle_t - \langle n(\mathbf{r})\rangle \langle n(\mathbf{r}')\rangle_t, \eeq
after a given time of flight $t$.
At zero temperature the density-density correlator is defined as follows:
\beq
\label{density_density}
\langle n(\mathbf{r}) n(\mathbf{r}')\rangle_t = \langle \Phi |U_0^\dag(t) \Psi^\dag(\mathbf{r})\Psi(\mathbf{r})\Psi^\dag(\mathbf{r}')\Psi(\mathbf{r}')U_0(t)|\Phi\rangle;
\eeq

\noindent where $|\Phi\rangle$ is the ground state inside the
trap, $U_0^\dag(t)$ is the free evolution operator and $\Psi$ is
the fermionic field. As
discussed in [\onlinecite{review08}], at sufficiently long times of flight there is a correspondence between the
density in the cloud and the momentum distribution inside the
trap, $\langle n(\mathbf{r}) \rangle_t \simeq \langle n_\mathbf{k}
\rangle_{\textrm{Trap}}$ with $\mathbf{k} = m\mathbf{r}/(\hbar
t)$. In the Mott phase, with one atom per site, a measure of the
density itself does not yield much information since it does not
allow to distinguish species with different spins.
However, one can extract interesting information about long-range
order in the trap from density-density correlation functions (see
Ref.[\onlinecite{altman04}] for a detailed calculation). In our
case these correlation functions would correspond to the spin-spin
correlation functions. One has indeed \beq \mathcal{G}(\mathbf{r}-
\mathbf{r}')=\mathcal{G}^1(\mathbf{r} -
\mathbf{r}')+\mathcal{G}^2(\mathbf{r} - \mathbf{r}'), \eeq with
\bea
\mathcal{G}^1(\mathbf{u}) &=& -\frac{1}{2}\frac{N}{W^2}\left( \frac{2\pi a_0}{a}\right)^2 \sum_\mathbf{G} \delta\left( \mathbf{u} + \frac{\hbar t}{m} \mathbf{G} \right), \\
\mathcal{G}^2(\mathbf{u} )&=& -2\sum_{i,j} e^{i\frac{m}{\hbar
t}(\mathbf{u}).(\mathbf{R}_i-\mathbf{R}_j)}\langle
\mathbf{S}_i.\mathbf{S}_j\rangle, \eea \noindent where $W = \hbar
t/(a_0 m)$, and $a_0$ is the width of the Wannier function on the
lattice,
$a$ is the lattice spacing and $\mathbf{G}$ is a vector of the
reciprocal lattice. The first term contains the usual Bragg peaks at reciprocal lattice wave vectors, while the second term is the static spin structure factor. Density correlations in the free-falling cloud can be a probe of spin correlations inside the trap.

\subsection{Result for a single ladder}

In order to compute $\mathcal{G}^2(\mathbf{r})$, we first resort
to the mapping of the ladder Hamiltonian onto the XXZ chain in the
large $J_\perp$ limit and write:
\bea
\chi(k_x,k_y)&=&\sum_{i,j} e^{i\mathbf{k}.(\mathbf{R}_i-\mathbf{R}_j)}\langle \mathbf{S}_i.\mathbf{S}_j\rangle \nn\\
&=& \frac{1}{2}(1 + \cos k_y a)\sum_{i,j}e^{ik_x(X_i-X_j)}\langle \mathcal{M}^z_i \mathcal{M}^z_j \rangle \nn \\
&+& (1 - \cos k_y a)\sum_{i,j}e^{ik_x(X_i-X_j)}\langle \sigma^+_i
\sigma^-_j \rangle,\nn \\
&=& \frac{1}{2}(1 + \cos k_y a)\chi_{zz}(k_x) + (1 - \cos k_y a)\chi_{+-}(k_x).\nn\\
\eea

\noindent where $i,j$ are the rung indices, $X_i$
is the position of rung $i$ along the longitudinal direction,
$\mathcal{M}_i^z= S^z_{i,1}+ S^z_{i,2}$ and the variables
$\sigma_i$ have been defined in Eq. (\ref{Spin_ladd_to_XXZ}).
It is worth
emphasizing that a $y$-dependent geometrical factor appears before
each sum, which means that for special values of $k_y$, i.e. special positions of observation inside the cloud, one can
preferentially observe rung-rung or transverse correlations.
The correlators $\langle \mathcal{M}^z_i\mathcal{M}^z_j \rangle$ and
$\langle \sigma^+_i \sigma^-_j \rangle$ can be computed at
zero temperature in the Luttinger liquid phase using
bosonization. While the correlators are easy to compute in the
Luttinger liquid phase, the calculation is much more
involved in the FIS. 
Generally one has:
\begin{widetext}
\beq
\langle \mathcal{M}^z(x) \mathcal{M}^z(0) \rangle = M^2 + \frac{1}{\pi^2}
\langle \partial_x \phi(x) \partial_x\phi(0) \rangle + \frac{2}{(2\pi
\alpha)^2} \cos (2 \pi M x) \langle \cos (2 \phi(x)) \cos (2 \phi(x)) \rangle ,
\eeq

\bea
\langle \sigma^+ (x) \sigma^- (0)\rangle = \frac{1}{\pi \alpha} \langle e^{i\theta(x)} e^{-i\theta(0)}\cos\left[2\phi(x) - \pi(2M-1)x\right]\cos\left[2\phi(0)\right]\rangle + \frac{\cos \pi x}{2 \pi \alpha} \langle
e^{i\theta(x)} e^{-i\theta(0)}\rangle.
\eea
\end{widetext}

Before computing explicitly those correlation functions, one remark can be made about the general form of $\chi(k_x,k_y)$. We have to compute sums of the form $ \sum_{i,j}e^{i\kappa(i-j)a}f(|i-j|)$, where $a$ is the lattice spacing and $i,j$ are integers running from $1$ to $N$, the length of a ladder. It is easily shown that:
\begin{widetext}
\beq
\sum_{i,j=1}^N e^{i\kappa(i-j)a}f(|i-j|) = N f(0) + 2\sum_{p=1}^{N-1} (N-p)\cos(p \kappa a)f(p).
\eeq
\end{widetext}

The maximum value  $N f(0) + 2\sum_p(N-p)f(p)$ of the sum is obtained for $\kappa \equiv 0[\frac{2\pi}{a}]$. Its minimum value is $N f(0) + 2\sum_p(N-p)(-1)^p f(p)$ and it is reached when $\kappa \equiv \frac{\pi}{a} [\frac{2\pi}{a}]$. The height of the peaks is partly controlled by the longitudinal size of the ladders.

\subsection{Correlation functions in the Luttinger liquid phase}

\begin{figure}
\epsfig{figure=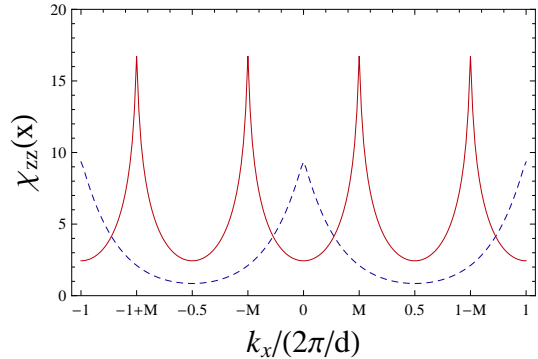,width=7cm} \caption{(Color online) Rung contribution in the spin part of the density correlations
(see Eq. (\ref{rung-rung})), in the Luttinger liquid phase. $k_x$ is in unit of $2\pi/d$. $\chi_{zz}$ in dimensionless.
We observe two sets of broad peaks: One at reciprocal lattice vectors, the other shifted by $\pm M$ (see text).
Here we have taken $M=1/4$, $K=0.6$ and a ladder with 100 rungs.} \label{Fig:XizzLL}
\end{figure}

\begin{figure}
\epsfig{figure=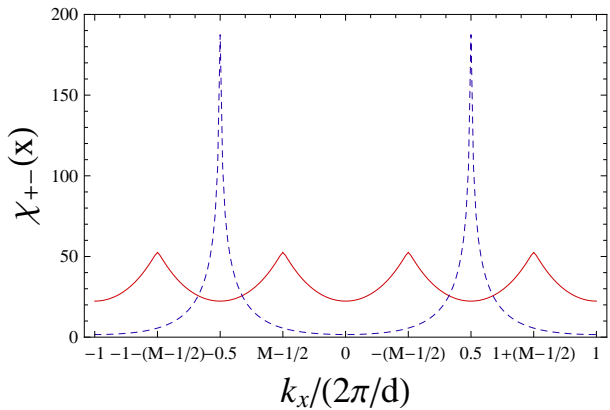,width=8cm} \caption{(Color online) Staggered contribution in the spin part of the density correlations (see Eq. (\ref{staggered})), in the Luttinger liquid phase. $k_x$ is in unit of $2\pi/d$. $\chi_{+-}$ in dimensionless. We also observe two sets of broad peaks. One is shifted by $\pm (M-1/2)$ with respect to reciprocal lattice vectors, the other is at half the reciprocal lattice vectors (see text). The latter is the most prominent, indicating quasi long-range antiferromagnetic order in the XY plane.}\label{Fig:Xi+-LL}
\end{figure}

As indicated earlier, in the Luttinger liquid phase, i.e. for $M \neq 1/2$ or $M = 1/2$ but weak frustration, all correlation functions decay as power laws, in accordance with the quasi-1D character of the ladders. The rung-rung correlation function reads:

\begin{widetext}




\beq \label{rung-rung} \langle \mathcal{M}^z(x) \mathcal{M}^z(0) \rangle = M^2 + \frac{K}{2\pi^2}
\frac{1}{x^2} + \frac{2}{(2\pi
\alpha)^2} \cos (2 \pi M x)\left( \frac{\alpha}{x} \right)^{2K},\eeq

As for the staggered correlation functions
(intrachain and interchain), they are embedded in:

\beq\label{staggered}
\langle \sigma^+ (x) \sigma^- (0)\rangle = \frac{1}{\pi \alpha} \cos\left(\pi(2M-1)x\right)\left( \frac{\alpha}{x}\right)^{2K+1/2K} + \frac{\cos \pi x}{2 \pi \alpha} \left( \frac{\alpha}{x}\right)^{1/2K}.
\eeq
\end{widetext}
The rung-rung correlation in equation (\ref{rung-rung}) contains
two type of terms: a Fermi liquid like mode around $q=0$, and another contribution at an
incommensurate mode with wave vector $q = 2\pi M$.
This mode decays as a power-law with a non-universal exponent which depends on the
interactions. The $XY$ correlations display power-law decays for
two modes, $q= \pi$ and $q = 2M-1$. Since $K>1/2$, the slowest
decaying mode is the antiferromagnetic one (i.e. $q=\pi$) in the $XY$
correlator. These properties are reflected through several peaks in $\chi_{zz}$ (Fig. \ref{Fig:XizzLL}) and $\chi_{+-}$ (Fig. \ref{Fig:Xi+-LL}). Due to the power-law decay of correlations these are broad peaks, as opposed to usual Bragg peaks appearing in $\mathcal{G}_1$. Most of them appear at incommensurate vectors, shifted from reciprocal lattice vectors by values depending on the magnetization $M$. The sharper and higher peaks appear in the staggered correlation function at half the reciprocal lattice vectors, indicating quasi-long range antiferromagnetic order in the $XY$ plane.\\

\begin{figure}
\epsfig{figure=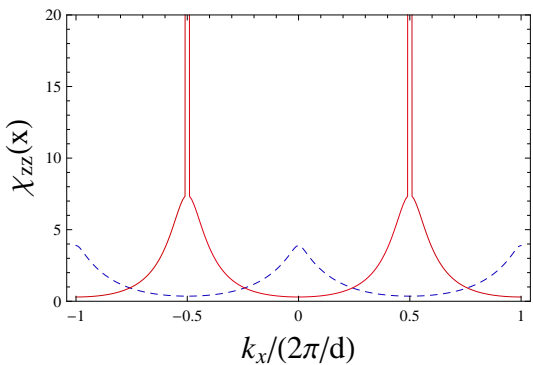,width=7cm} \caption{(Color online) Rung contribution in the spin part of the density correlations, in the FIS. $k_x$ is in unit of $2\pi/d$. Former incommensurate peaks coalesce into Bragg peaks at half-reciprocal lattice vectors. Former peaks at reciprocal lattice vectors are rounded. Here $M=1/2$, $K=0.25$} \label{Fig:Xizz_plateau}
\end{figure}

\begin{figure}
\epsfig{figure=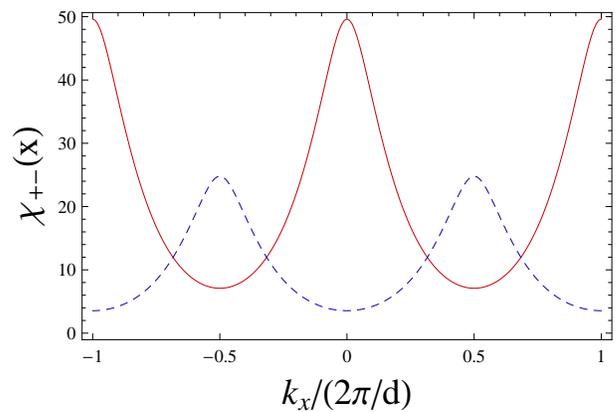,width=8cm} \caption{(Color online) Staggered contribution in the spin part of the density correlations, in the FIS. $k_x$ is in unit of $2\pi/d$. Incommensurate peaks coalesce to reciprocal lattice vectors. All peaks get rounded by the exponential damping of the modes. }\label{Fig:Xi+-_plateau}
\end{figure}

\subsection{Correlation functions in the fully incompressible state}
\label{app:fully-incompressible} In the FIS, the system opens a spin gap, and
excitations are now described by a sine-Gordon (sG) Hamiltonian:
\bea
H_{boson} &=& \frac{1}{2\pi}\int dx \left[ uK\left( \partial_x \theta(x)\right)^2 + \frac{u}{K} \left( \partial_x \phi(x)\right)^2 \right] \nn \\
&-& \frac{2g_3}{(2\pi \alpha)^2} \int dx \cos \left(4 \phi(x)\right).
\eea
Every excitation of the sG model is massive and will
therefore lead to exponentially decaying correlations.  To gain some insight on the precise form of correlators, let us
remember that the sine-Gordon Hamiltonian is obtained from the $XXZ$ chain Hamiltonian

\beq
H_{XXZ}= \sum_{i=1}^N \left[ J_{xy} \left( \sigma^x_i \sigma^x_{i+1} + \sigma^y_i \sigma^y_{i+1}\right) + J_z\sigma^z_i \sigma^z_{i+1}\right].
\eeq

\noindent Since the $J_{xy}$ interaction is frozen on the plateau, the physics is that of the Ising model in $1+1$ dimension, away from criticality. One can map the latter model to free Majorana fermions, and correlations can be computed exactly.
A theory of free Majorana fermions is obtained from the sine-Gordon model, for $K=1/4$. One intuitive way to look at the problem is to use the Luther-Emery trick\cite{luther} and make a change of variable $\tilde{\phi} \rightarrow \phi/\sqrt{K}$, that transforms the cosine operator into $\cos 2\tilde{\phi}$, which is a simple backscattering term in the fermion language. The resulting refermionized Hamiltonian is that of two species of fermions, right-moving and left-moving, with back-scattering -- that in turn can be written in terms of free Majorana fermions. Using this trick one is able to compute $\langle e^{i\theta(x)} e^{-i\theta(0)}\rangle$ exactly:

\bea \langle e^{i\theta(x,\tau)}e^{-i\theta(0,0)}\rangle &=& \left(
\frac{ \Delta}{u}\right)^2 \left[ K_1\left(
\frac{\Delta}{u}\sqrt{x^2 + \alpha^2} \right)^2 \right.\nn \\
 &-& \left. K_0\left( \frac{\Delta}{u}\sqrt{x^2 + \alpha^2}
\right)^2 \right],  \eea

\noindent where $K_0$ and $K_1$ are modified Bessel functions of the second kind. The short distance behavior of this function is simply $(x^2 + \alpha^2)^{-1}$. This coincides with the Luttinger liquid result, as $K=1/4$ on the plateau. However at large distances, it decays exponentially.
In $\chi_{+-}$, former sharp peaks at half reciprocal lattice vectors are greatly reduced and loose their cusp while incommensurate peaks coalesce at reciprocal lattice vectors, losing their cusp as well (see Fig. \ref{Fig:Xi+-_plateau}).
In order  to get $\langle \cos (2 \phi(x)) \cos (2 \phi(0)) \rangle$, we use the same refermionization procedure
and mapping to the 2D Ising model off criticality (see [\onlinecite{tsvelik}], chap. 18, for details)
and obtain
\beq
\label{corr_plateau}
\langle \cos 2\phi(x) \cos 2\phi(0) \rangle \simeq \left(1 + 2K^2 K_0\left(\frac{\Delta}{u}\sqrt{x^2 + \alpha^2}\right)^2\right).
\eeq

\vspace{0.5cm}
We thus find that the function $\langle \cos (2 \phi(x)) \cos (2 \phi(0)) \rangle$ decays exponentially to 1.
As a consequence the former incommensurate broad peaks in $\chi_{zz}$ coalesce into sharp Bragg peaks at half reciprocal lattice wave vectors (see Fig. \ref{Fig:Xizz_plateau}).
Using the following formula
\beq \langle \mathcal{T}_\tau \p_x
\phi(x,\tau) \p_x \phi(0,0) \rangle = \lim_{\eta \rightarrow 0}
\frac{1}{\eta^2}\p_x^2 \langle
e^{i\eta\phi(x,\tau)}e^{-i\eta\phi(0,0)}\rangle, \eeq
one explicitly finds at zero temperature:
\bea
\langle \p_x \phi(x) \p_x \phi(0) \rangle &=&
\frac{K\Delta^2}{4u^2}\left[ K_0\left( \frac{\Delta}{u}\sqrt{x^2 + \alpha^2}\right) \right. \nn \\
&+& \left. K_2\left( \frac{\Delta}{u}
\sqrt{x^2 + \alpha^2}\right)\right]. \eea
At short distances $\langle \partial_x \phi(x) \partial_x\phi(0) \rangle$ behaves as $(x^2+\alpha^2)^{-1}$ just as in the Luttinger liquid phase, but it is exponentially suppressed at large distances. In $\chi_{zz}$ former peaks at reciprocal lattice vectors change shape by loosing their cusp (Fig. \ref{Fig:Xizz_plateau}). To summarize, the appearance of Bragg peaks at half reciprocal lattice vectors in $\chi_{zz}$ and the clear damping of these modes in  $\chi_{+-}$  indicates antiferromagnetic ordering in the $z$ direction which corresponds  in that case to
an alternation of singlets and triplets on the rungs of a ladder.


\newpage
\section{Conclusions}\label{sec:conclusion}
In summary, frustrated spin ladders can be simulated by an optical
superlattice. We have shown that a phase transition from the spin liquid to a fully
incompressible phase or magnetically ordered phase is possibly
realized through adjusting the lattice parameters controlled by
laser beams. 
The spin-gapped state is expected to occur at sufficiently low temperatures $T$, less than the gap $\Delta$. From the study of section \ref{sec:4} it appears that $\Delta$ is of the order of $J_1$. Therefore, the observation of this state demands the same kind of experimental challenge on temperature as the realization of the Néel phase in an optical lattice.
Provided temperature is low enough, the coalescence of incommensurate peaks in the staggered and longitudinal density-density correlation functions into Bragg peaks at 
half-reciprocal or reciprocal lattice vectors would provide a concrete signature for the observation of the transition.
 Our results show that strongly correlated cold atoms
in optical lattices provide a route to observe this quantum phase
transition. Manipulating the amplitudes and wave vectors of laser
beams, the coupling between spins can be adjusted in a wide range.
Recent developments of controlling the cold atoms in optical
lattices allow for the experimental investigation of our
prediction in the future.\\

{\bf Acknowledgements:} We would like to acknowledge fruitful discussions with N. Laflorencie.


\begin{thebibliography}{30}
\bibitem{note} Let us also note that while
rigorously the potential (\ref{eq:potential}) would have $t'_2=0$,
we keep it for generalities in our model Hamiltonian.
\end{thebibliography}
\end{document}